\documentclass[letterpaper,final]{appolb}
 \usepackage{epsfig}
\begin{document}

\newcommand{\be}{\begin{eqnarray}}
\newcommand{\ee}{\end{eqnarray}}
\newcommand\del{\partial} 
\newcommand\nn{\nonumber}
\newcommand{\mat}{\left ( \begin{array}{cc}}
\newcommand{\emat}{\end{array} \right )}
\newcommand{\vect}{\left ( \begin{array}{c}}
\newcommand{\evect}{\end{array} \right )}
\newcommand{\tr}{\rm Tr}

\title{Triage of the Sign Problem%
\thanks{We thank the organizers of the workshop
 ``Random Matrix Theory: From fundamental physics to applications'' 
for their generous hospitality.} }%

\author{
 K. Splittorff
\address{
Niels Bohr Institute, Blegdamsvej 17, DK-2100, Copenhagen {\O}, 
Denmark} 
\and
J.J.M. Verbaarschot
\address{
Niels Bohr International Academy, Blegdamsvej 17, DK-2100, Copenhagen {\O},
Denmark,} 
\address{Niels Bohr Institute, Blegdamsvej 17, DK-2100, Copenhagen {\O},
  Denmark,} 
\address{Department of Physics and Astronomy, SUNY, Stony Brook,
New York 11794, USA}
}

\maketitle
\begin{abstract}
We discuss the sign problem in QCD at nonzero chemical potential
and its relation with chiral symmetry breaking and the spectrum
of the Dirac operator using the framework of chiral random matrix theory.
We show that the Banks-Casher formula is not valid
for theories with a sign problem and has to be replaced by
an alternative mechanism that is worked out in
detail for QCD in one dimension at nonzero chemical potential.

\end{abstract}

\section{Introduction}
Despite of tremendous efforts \cite{schmidt,deforcrand,misharev,split,cm}, the
phase diagram of QCD in the chemical potential temperature plane
is only known for $\mu = 0$ and at asymptotically large densities.
In between these extreme limits,
the only firm result at low temperatures is the transition to
nuclear matter when the sum of the chemical potential and the
binding energy of nuclear matter is equal to the nucleon mass.
Most results at intermediate values of the chemical potential are
based on models such as the Nambu-Jona-Lassinio model \cite{rav,pnjl}, 
strong coupling expansions \cite{damgaard,nish}, ADS-CFT dualities \cite{kim,bergman,parnachev}
and random matrix theory \cite{shrock,janikcri}.
 
The reason for the lack of first principle calculations is the 
phase of the fermion determinant at nonzero chemical potential which
invalidates probabilistic methods. When the sign problem is
mild, though, simulations can be performed by absorbing the phase in 
the observable which is known as reweighting.
The real merit of the work by Fodor and Katz \cite{fodor} is the realization
that the sign problem is much less severe close to $T_c$ so that 
sophisticated reweighting methods have a chance to work. In fact, results
close to $T_c$ and small chemical potential obtained with a variety of 
methods
such as imaginary chemical potential \cite{maria,owe1}, 
Taylor expansion \cite{Allton1,gupta}, reweighting \cite{fodor},
the canonical ensemble \cite{AFHL}  and
the density of state method \cite{schmidtDOS,azcoiti}, are in close agreement.

In this lecture we will discuss the average phase factor of the fermion
determinant. 
It provides a direct measure for the severity of the
sign problem. 
We will
focus on  the average phase factor in the microscopic domain of QCD where
{\it nonperturbative} analytical results can be obtained by exploiting the
equivalence of QCD and chiral random matrix theory
\cite{SVprl2,SVnuc}.
{\it Perturbatively}, the average phase factor can be calculated
by means of chiral perturbation theory up to temperatures close to $T_c$
and chemical potentials up to the pion mass \cite{SplittVerbnew}.
 We will
find that the sign problem is necessarily severe when $\mu > m_\pi/2$.
The reason is that
exponentially large contributions to the partition function
have to be canceled in order to obtain a chiral condensate that 
has sensible physical properties such as a discontinuity when the
quark mass crosses the imaginary axis. 

One disturbing observation is that the relation between 
the spectral density of the Dirac operator and the discontinuity of 
the chiral condensate seems to be violated for QCD at nonzero 
chemical potential. From an analysis in the microscopic
domain of QCD 
it was found  \cite{OSV} that the discontinuity arises 
due to an alternative mechanism. Lending support to
its universality, the same mechanism is at work
\cite{lorenzo} for QCD in one dimension. 
This case, because of its simplicity, will
be discussed in detail below.

In this
lecture we will show that chiral random matrix theory has added significantly
to our understanding of chiral symmetry breaking  and the sign problem
for QCD at nonzero chemical potential. This adds to a long list of 
successes of random matrix theory in this field such as the
understanding of quenched approximation \cite{misha}, the critical endpoint
in QCD \cite{shrock,janikcri}, the macroscopic 
spectral density \cite{misha,janik,janik1,trimu,janik2,jarosz,TV}, the
microscopic spectral density 
\cite{SplitVerb2,O,Akemann-Wettig,AOSV,bloch,dam,gerbit} and 
Yang-Lee zeros \cite{misha,adam,misha2}. For more
successes we refer to reviews of this subject \cite{tilorev,gerrev}.

After some introductory remarks on QCD at nonzero chemical potential in section
2, we will discuss the sign problem in section 3. The microscopic domain
of QCD will be introduced in section 4. Results for the average phase factor
will be presented in section 5,  and their relation with the Dirac spectrum
is examined in section 6. Section 7 contains a detailed discussion of the
alternative to the Banks-Casher formula for the example of one-dimensional
QCD at nonzero chemical potential. Concluding remarks are made in section 8.

\section{QCD at Nonzero Chemical Potential}

The QCD partition at temperature $ 1/\beta$ and chemical potential
$\mu$ is given by 
\be
Z_{\rm QCD} &=&\sum_k e^{-\beta ( E_k -\mu)} ,
\ee\noindent
where the sum is over all states. 
This partition function can be rewritten as a Euclidean quantum field theory
\be
 Z_{\rm QCD} &=&\langle \prod_f\det({D} +  m_f
+ {\mu} \gamma_0)\rangle_{\rm YM}, 
\ee
where the average is over the Yang-Mills action. The Dirac operator is
denoted by $D$ and the product is over $N_f$ flavors with mass $m_f$.
The Dirac operator is nonhermitean whereas $\mu \gamma_0$ is Hermitean
so that the total Dirac operator has no hermiticity properties.
In lattice QCD, the chemical potential enters on each time-like link such that
forward hopping is enhanced
by $\exp(\mu)$ whereas backward hopping is suppressed 
by $\exp(-\mu)$ \cite{HK}.
The asymmetry between forward and backward propagation also implies that
the Dirac operator is nonhermitean for  $\mu \ne 0$:

\be
\det(D+m+\mu\gamma_0) = |\det(D+m+ \mu \gamma_0)| e^{i\theta} = 
\prod_k(\lambda_k +m). 
%\qquad \theta \ne 0.
\ee
If the average phase factor vanishes in the thermodynamic limit, Monte
Carlo simulations are not possible. This problem is known as the
sign problem.
Nevertheless,  we emphasize that it is our aim  to understand 
QCD at {$\mu \ne 0$} starting from
{\it first} principles.

\begin{figure}[!t]
\centerline{\includegraphics[width=10cm]{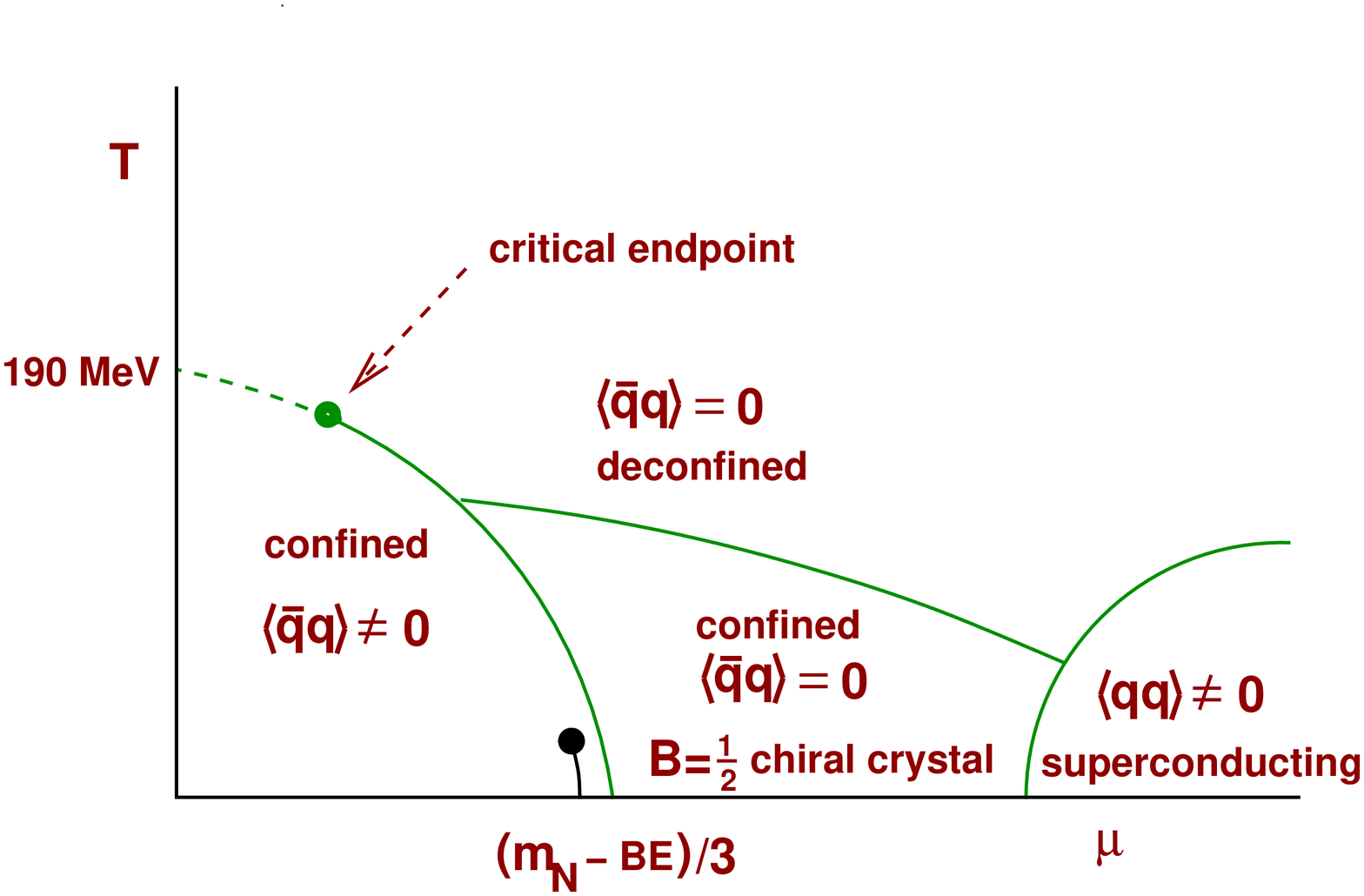}}
\caption{A schematic phase diagram of QCD in the temperature chemical
potential plane}
\label{phdia}
\end{figure}

Let us now discuss the phase diagram in the chemical potential temperature
plane. A schematic phase diagram is shown in Fig. \ref{phdia}. Even though
the $\mu = 0$ axis is rather well-understood, there is still an ongoing
controversy about the value of the crossover temperature. Whether its
value of 190 MeV \cite{karschcol} or 170 MeV \cite{fodort} 
or somewhere in between, 
will definitely
be resolved by future lattice simulations.
At asymptotically large values of the chemical potential, 
perturbative calculations, show that QCD
is superconducting \cite{son}. We will not further discuss this region.

At low temperature and  $ \mu < (m_N - BE)/3$ with $ BE$ the binding energy
per nucleon of nuclear matter, only the vacuum state contributes to
the QCD partition function, and its free energy is $\mu$-independent.
At $\mu= (m_N-BE)/3$ a transition to nuclear matter takes place. These 
are the only solid results at intermediate densities. All other results
at intermediate density are model dependent. We agree with
McLerran and Pisarki \cite{larry} that confining forces play an essential
role, and that one better relies on models where confinement is manifest.
One such model is the Skyrme model which is believed to be an accurate
description of QCD at large $N_c$. 
With increasing density this model
undergoes a transition \cite{klebanov} to a qualitatively different phase. 
It was
shown in \cite{jacksonv,castil} that the  
dense phase that minimizes the energy 
is a chiral crystal of $B = \frac 12$ objects
\cite{mantongold} which has restored chiral symmetry. This state is
strongly bound with a binding energy of at least 100 MeV per nucleon over
nuclear matter. We expect that it takes similar temperatures to melt
this phase.  
Most recently such scenario was advocated in 
\cite{larry} for the large $N_c$ phase diagram of QCD.
If indeed the
phase transition to the chiral symmetric phase is of first order,
the critical line 
should end in a critical endpoint to allow for a crossover transition
at small chemical potential.

\section{Triage of the Sign Problem}

A quantitative measure for the severity of the sign problem 
is given by the ratio of the QCD partition function and
the phase quenched QCD partition function,
 \be 
\frac{{\langle {\det}^2(D+m +\mu\gamma_0)\rangle}}
{ {\langle |\det (D+m +\mu\gamma_0)|^2 \rangle}}
\sim e^{-V(F_{N_f=2} -F_{\rm pq})}.
\label{rat}
\ee
Because both the numerator and denominator are physical partition functions
they can be expressed in terms of an extensive free energy. At nonzero
temperature, the difference $F_{N_f=2} -F_{\rm pq}$ is always nonzero so that, 
in the thermodynamic limit, the sign problem becomes prohibitively severe.
There is no reason, though, to doubt that, 
despite these cancellations, the large volume limit
is still smooth. What happens is that, with increasing volume, it becomes
more and more difficult to generate the QCD partition function from
the phase quenched ensemble. 
By factoring the determinant of the Dirac operator into its absolute and
phase factor, $\exp( i\theta)$, the ratio (\ref{rat}) can be interpreted
as the phase quenched expectation value of $\exp(2i\theta)$,
\be
\frac{{\langle {\det}^2(D+m +\mu\gamma_0)\rangle}}
{ {\langle |\det (D+m +\mu\gamma_0)|^2 \rangle}}
= \langle e^{2i\theta} \rangle_{\rm pq}.
\ee

The phase quenched QCD partition function can be rewritten as the
expectation value
\be
\langle |\det (D+m +\mu\gamma_0)|^2 \rangle  =
\langle\det (D+m +\mu\gamma_0) \det (D+m -\mu\gamma_0)\rangle.
\ee
This means that the two flavors have opposite charge with respect
to the chemical potential. In other words, $\mu$ can be interpreted
as an isospin chemical potential. Therefore, in the low-temperature
limit, the free energy of phase quenched QCD remains constant for
$\mu < m_\pi/2$. At $\mu = m_\pi/2$ a phase transition to a Bose-condensed
phase of pions takes place so that the free energy becomes $\mu$-dependent
for $\mu >m_\pi/2$ \cite{KSTVZ,SS,KS}. 
In the low temperature limit, the free energy of QCD is $\mu$-independent
for $\mu < (m_N-BE)/3$. This implies that the free energy of QCD and
phase quenched QCD are different for $m_\pi/2 <\mu < (m_N-BE)/3$ resulting in
a severe sign problem.

Below we will analyze the average phase factor in the microscopic domain of QCD
and for one-dimensional QCD.

\section{ Microscopic Domain of QCD and Random Matrix Theory}

The QCD Dirac spectrum can be probed by including
additional bosonic and fermionic determinants in the partition
function with quark masses equal to complex parameters that
can be varied independent of the QCD quark masses. For definiteness
lets us assume that we have $N_f$ original quarks and $s$ additional
quarks with masses  given by
\be
m_1, \cdots m_{N_f}\quad  {\rm and} \quad z_1, \cdots z_s, 
\ee
respectively.
For fixed QCD quark masses,
it is always
possible to choose these additional quark masses such that the
associated Goldstone bosons are much lighter than the Goldstone
bosons of the original quarks (the case of massless quarks has to
be treated separately; we assume that all quark masses are nonzero). 
Using the same arguments as
given by Gasser and Leutwyler for the $\epsilon$-domain of QCD \cite{GL},
to leading order in the chiral expansion of the $z_k$-quarks, the
partition function factorizes as \cite {Vplb}
\be
Z(m_1,\cdots,m_{N_f}, z_1, \cdots, z_s) =
Z_{QCD}(m_1,\cdots,m_{N_f})  
Z(z_1, \cdots, z_s) 
\ee
if the Compton wave lengths of the additional bosons containing the
quark masses $z_k$ are much larger than the size of the box. 
Using the Gell-Mann-Oakes-Renner relation, 
the {\it microscopic} domain of QCD is given by \cite{Vplb}
\be
|z_k| \ll \frac {F^2}{\Sigma \sqrt V},
\label{validity}
\ee
where $\Sigma $ is the chiral condensate, $F$ the pion decay constant and
$V$ the volume of the box. 
In solid state physics, this scale
for $z$ is known as the Thouless energy and was discussed in the 
context of QCD in \cite{janikdis,james,guhr}.
Of course, we can consider QCD with quark masses $m_k$ in the microscopic
domain (\ref{validity}) which is also known as the $\epsilon$-domain of
QCD. This domain will also be called the microscopic domain of QCD. 

In the domain (\ref{validity}), the $z_k$ dependent part of
the partition function is given by a unitary matrix integral. At 
$\mu\ne 0 $ invariance arguments lead to the following partition function
of the following form \cite{KSTVZ,TV}
\be 
Z(z_1,\cdots, z_s) = \int dU {\rm Sdet}^\nu U
e^{V\Sigma[{\rm Str} (MU^\dagger +M^\dagger U)]
-\frac 14 F^2 V {\rm Str} [B,U] [B, U^\dagger]},
\label{zgold}
\ee
with quark mass matrix given by $M= {\rm diag}(z_1,\cdots, z_s)$ and charge
matrix equal to $B={\rm diag}(q_1,\cdots, q_s)$. The superdeterminant and
the supertrace are denoted by Sdet and Str, respectively.
If we have $f$ fermionic
quarks and $s-f$ bosonic quarks, the integral is over a supergroup
with bosonic sector equal to the product of  $U(f)$ and the positive
definite matrices $Gl(s-f)/U(s-f)$. For convergence reasons, the quark
mass matrix involving the bosonic quarks has to be properly adjusted
\cite{SplitVerb2} and additional conjugate quarks may have to be introduced
\cite{SVbos}. 

Another representation of the  partition function (\ref{zgold}) is
the large-$N$ limit of a
random matrix model with the symmetries of $Z(z_1,\cdots,z_s)$.  
This random matrix model is obtained by 
replacing the matrix elements of the Dirac operator
by Gaussian random numbers \cite{SV,V},
\be
D = \mat m & iW+\mu \\ i W^\dagger +\mu& m \emat, \quad P(W) \sim 
e^{-N {\rm Tr} W^\dagger  W},
\ee 
where $W$ is in general an $N \times (N+\nu)$ matrix  with $\nu$ the
topological charge. As was shown in \cite{damuni}, the properties of
this theory in the microscopic domain are not sensitive to the details
of the probability distribution. The reason is that for $N\to \infty$
with $m N$ fixed, the random matrix model has a mass gap, so that it
becomes a theory of Goldstone bosons dictated by the pattern of 
spontaneous symmetry breaking with partition function given
by (\ref{zgold}).

Philosophically, this is important because of the realization that 
chaotic motion dominates the dynamics of quarks at low energy.
Practically, this is useful because it enables us to
use powerful random matrix techniques to calculate physical
observables.

\section{The Phase of the Fermion Determinant}

Let us now calculate the average phase factor
in the microscopic domain
of QCD. For simplicity we consider the quenched case so that
\be
\langle e^{2i\theta} \rangle = \langle \frac {\det(D+m+\mu\gamma_0)}
{\det(D^\dagger+m+\mu\gamma_0)}\rangle.
\label{thth}
\ee
Since the average phase factor is the ratio of two
 partition functions,
its value is necessarily real and nonnegative. We evaluate the ratio
(\ref{thth}) in the microscopic domain of QCD
where only the contribution of the zero momentum Goldstone modes has 
to be taken into account.
The partition function (\ref{thth}) 
has four different Goldstone modes. If we denote the fermionic
quark by $f$ and the bosonic quark by $b$, they are given by
 \be
\bar f f,\quad \bar b b, \quad \bar b f, \quad \bar f b.
\ee
The first two are neutral, but because the charge of conjugate quarks
is opposite to that of regular quarks, the last two have charge 
$\pm 2$. The masses
of the corresponding Goldstone bosons are thus given by
 \be
m_{\bar f f}= m_\pi,\quad m_{\bar b b}= m_\pi, 
\quad m_{\bar b f} = m_\pi + 2\mu, \quad m_{\bar f b}= m_\pi - 2\mu.
\ee
When $\mu < m_\pi/2$ only the vacuum state contributes to the partition
function so that the free energy is
a $\mu$-independent constant and 
the average phase factor is given by the product of the square root of the 
curvatures 
of the Goldstone modes,
\be
\langle e^{2i\theta}\rangle = 
\frac{(m_\pi -2\mu)(m_\pi+2\mu)}{m_\pi^2} .
\ee
For $\mu>m_\pi/2$ the massless pion Bose condenses, and the free energy
becomes $\mu$-dependent so that the average phase factor becomes zero
in the thermodynamic limit. We thus find
\be
\langle e^{2i\theta} \rangle = \theta(m_\pi -2 |\mu|) 
\left(1- \frac{4\mu^2}{m_\pi^2}\right).
\ee
\begin{center}
\begin{figure}[!t]
{\includegraphics[width=6.2cm,clip=]{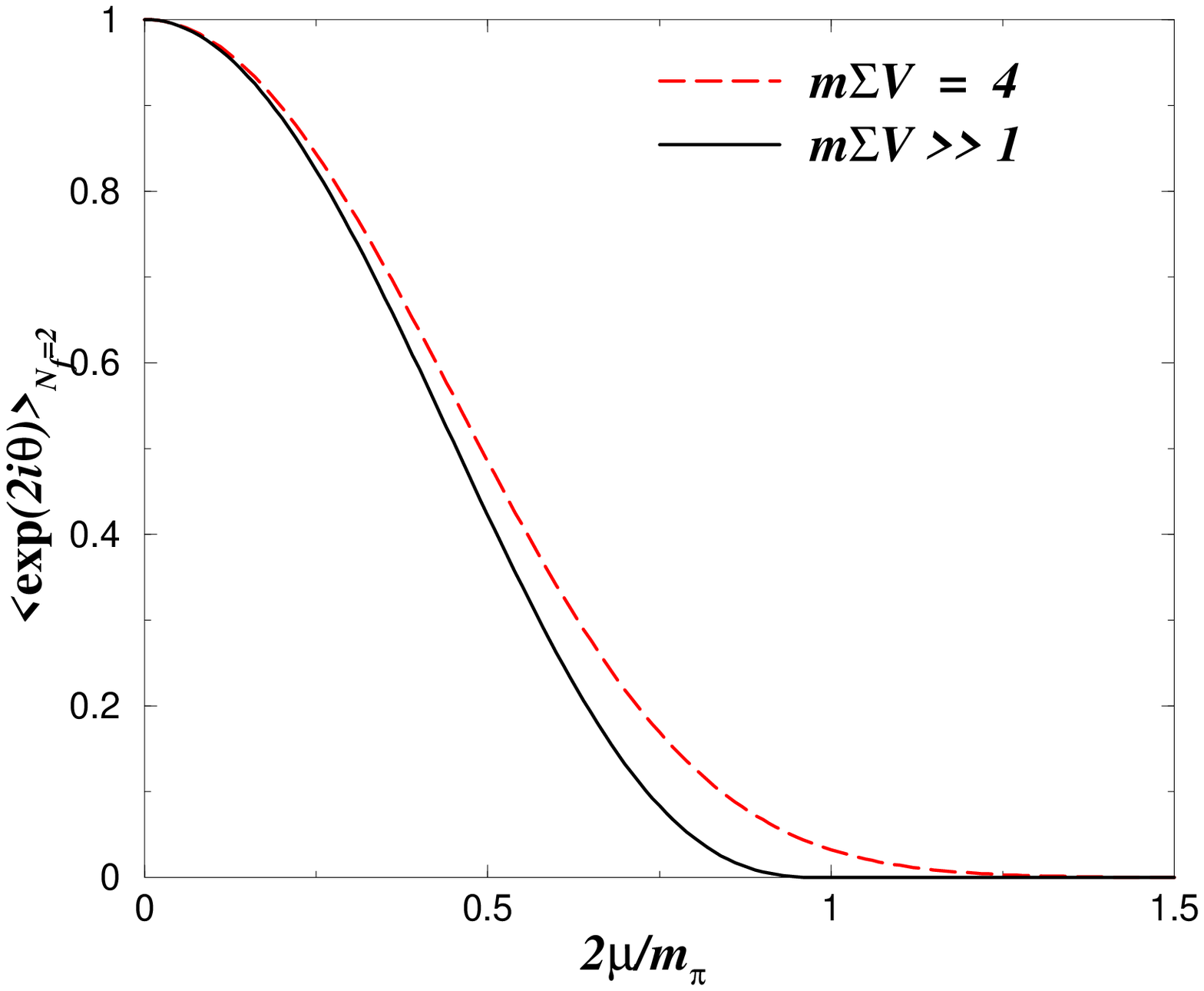}}
\includegraphics[width=6.2cm,clip=]{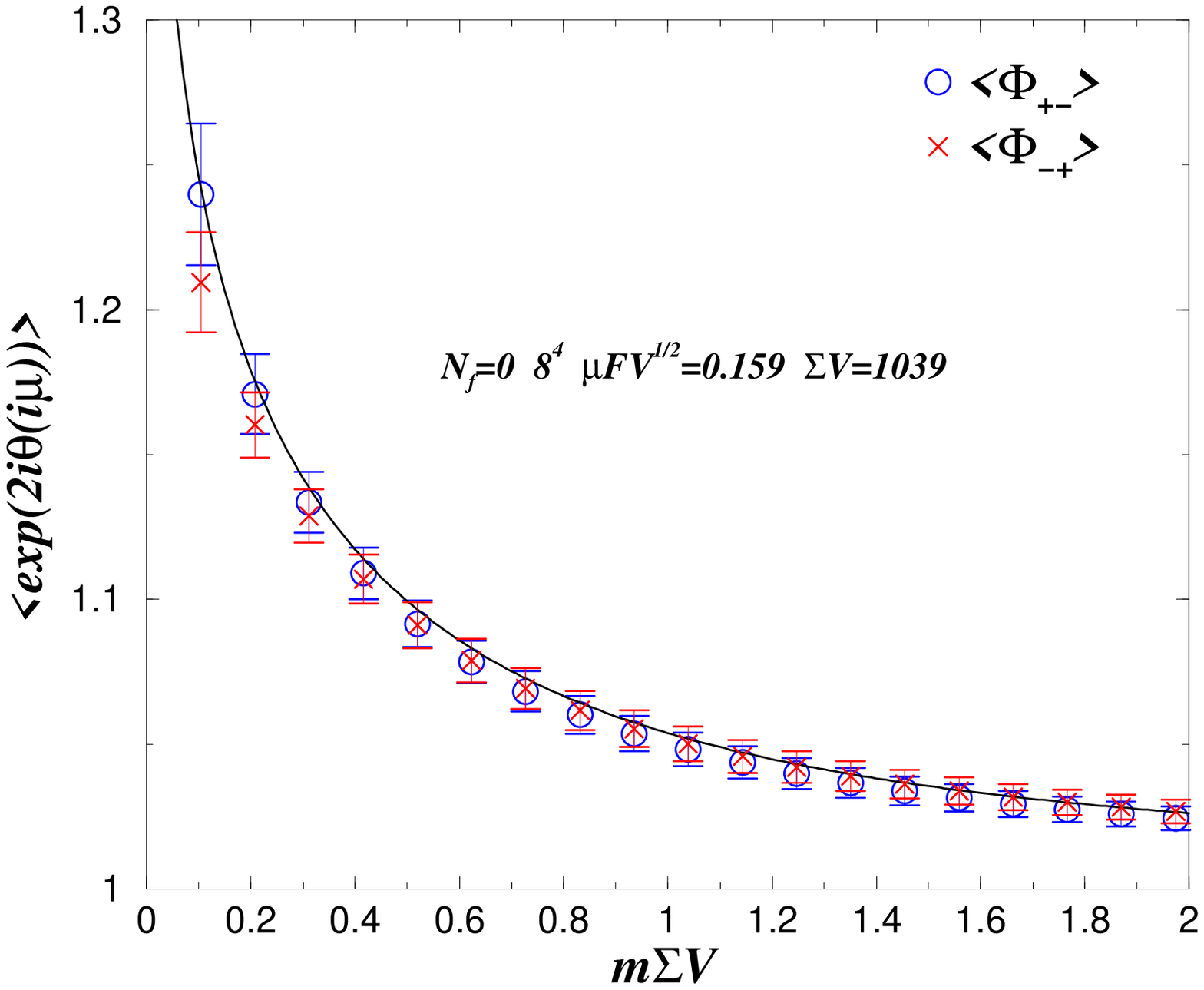}
\caption{The microscopic result of the average phase factor (left)
and the average phase factor for imaginary chemical potential (right).
The points with error bars in the right figure are lattice results
obtained in \cite{BenKim}.}
\label{figphase}
\end{figure}
\end{center}

In \cite{SVprl2} the average phase factor was calculated by means of the
complex orthogonal polynomial method  of \cite{O,AP}. The result is given by
\be\label{phex}
&&\langle e^{2i\theta} \rangle_{N_f = 0}= 1- 
4\hat \mu^2 I_0(\hat m) K_0(\hat m)  \\
&& -\frac{1}{4\hat{\mu}^2} e^{-2\hat{\mu}^2-\frac{\hat{m}^2}{8\hat\mu^2}} 
\int_{\hat{m}}^\infty dx x e^{ -\frac {x^2}{4\hat\mu^2}}
K_0\left ( \frac{x\hat{m}}{4\hat\mu^2}\right ) \left(I_0(x)\hat{m}
I_1(\hat{m})-x I_1(x)I_0(\hat{m})\right).\nn
\ee
The average phase factor can also be calculated using an imaginary chemical
potential \cite{BenKim}. 
However, this gives only the first two terms and misses the 
non-analytic term in the above expression. The non-analytic term is 
essential when $\mu$ approaches $m_\pi/2$ and cancels the analytic term for
$\mu > m_\pi/2$.

In Figure \ref{figphase} we show the analytical result for the average
phase factor weighted with the determinant for two flavors (left figure). 
The solid curve is the mean field result. In the right figure we
compare the quenched average ``phase factor'' for imaginary chemical potential
with lattice simulations on an $8^4$ lattice (see \cite{BenKim}). 
The solid curve is
given by the analytical continuation ($\mu^2 \to -\mu^2$) of the first
two terms in (\ref{phex}).

\section{Average Phase Factor and Dirac Spectrum}

\begin{figure}[!t]
\includegraphics[width=5.5cm]{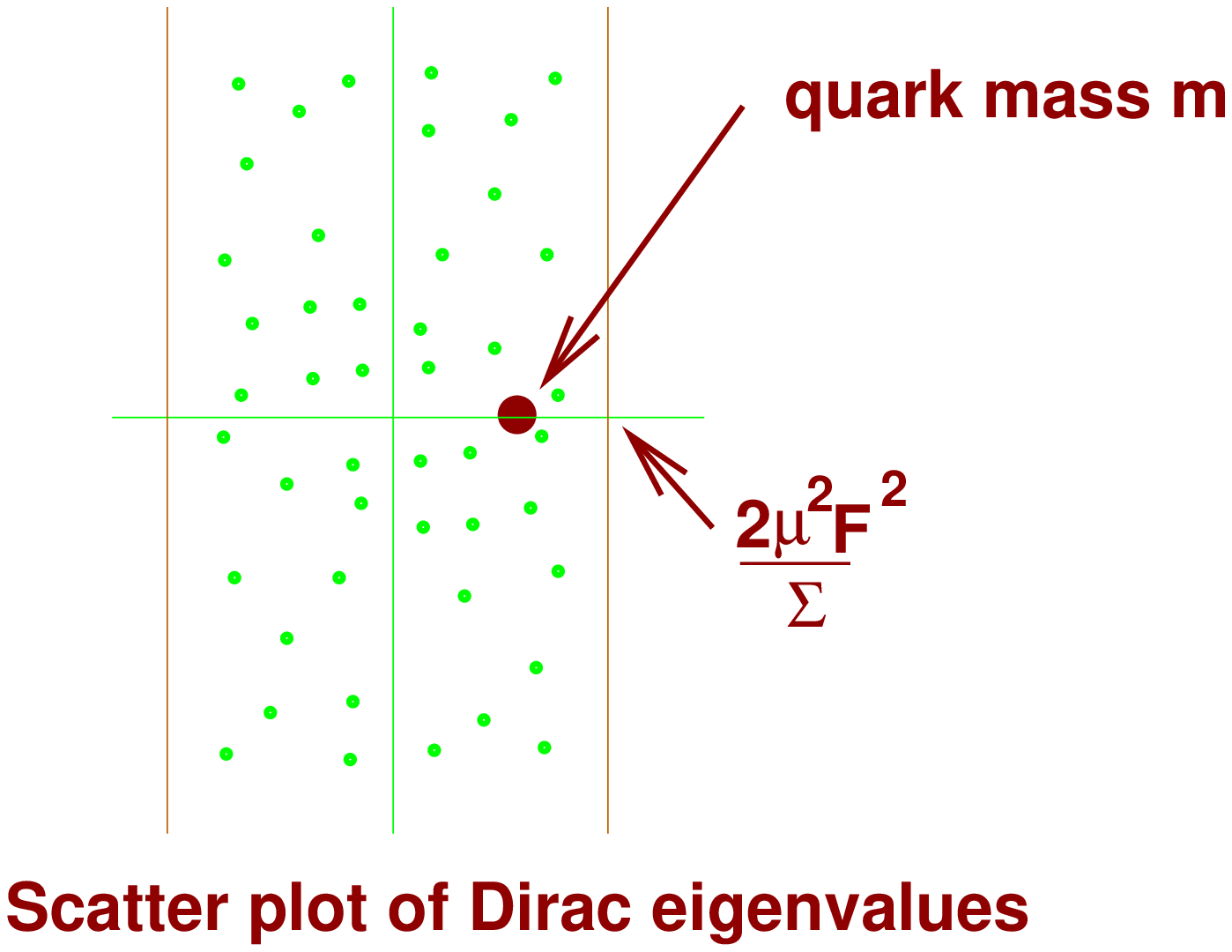}\hspace*{0.2cm}
\includegraphics[width=6.5cm]{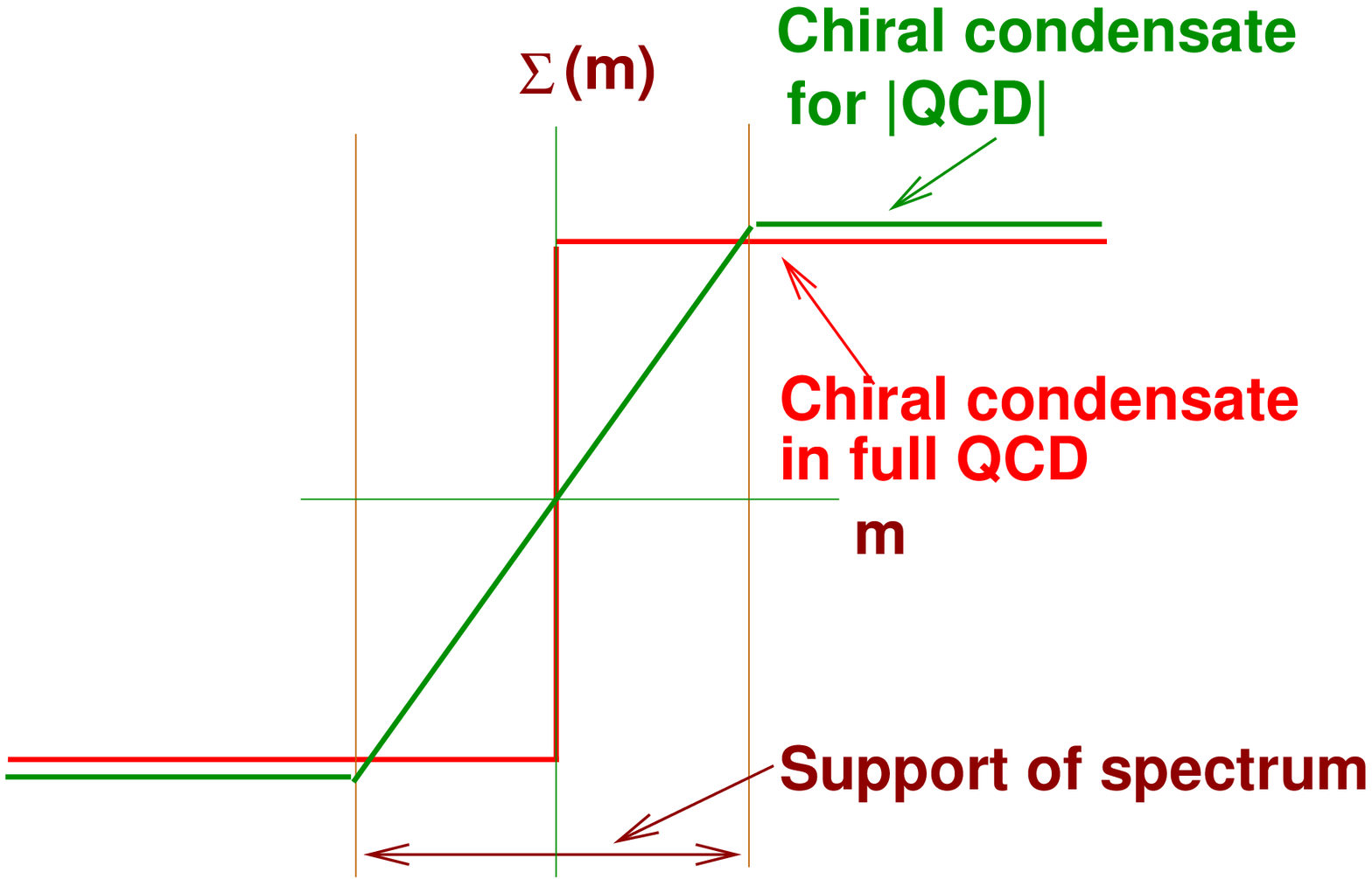}
\caption{Scatter plot of Dirac eigenvalues and $\mu \ne 0$ (left), and
in the right panel, we show
the mass dependence the chiral condensate for full QCD (red curve)
and quenched or phase quenched QCD (green curve).}
\label{specmu}
\end{figure}

The sign problem becomes inevitable when the quark mass is inside the domain
of eigenvalues. This should be obvious because for eigenvalues close to
the mass, small variations of the gauge field result in large
phase fluctuations. In figure \ref{specmu} we illustrate the distribution
of the Dirac eigenvalues.
The width of the spectrum
can be obtained from chiral perturbation theory
\cite{TV} or chiral random matrix theory  \cite{misha}.
In the quenched case or the phase quenched case the chiral condensate
at $m$ can be interpreted as the planar electric field at $m$ of charges
located at the position of the eigenvalues. Elementary electrostatics
dictates that the chiral condensate behaves as the green curve in 
the right panel of Figure \ref{specmu}. In particular, there is no
discontinuity at $m = 0$. Because the low temperature limit of the 
free energy of full QCD does not depend on the chemical potential, the
chiral condensate of full QCD should have a discontinuity at $m =0$ for 
$\mu < m_N /3$. What has to happen is that the phase of the fermion
determinant has to cancel the decrease in free energy that takes 
place in the quenched or phase quenched theory for $ \mu > m_\pi/2$.
Therefore, we necessarily have a severe sign problem in this domain.

The puzzle that the chiral condensate remains the same as the result
of strong cancellations is known as the ``Silver Blaze Problem''
\cite{cohen}. This name has been  inspired by the title of
a Sherlock Holmes novel by Conrad Doyle. The analogy is that the dog
did not bark although  the racing horse ``Silver Blaze'' disappeared. 

In the literature it has been stated that average spectral density
of full QCD will show an accumulation of spectral density on the imaginary
axis consistent with the Banks-Casher formula. 
We now understand that this is not the case.
The average spectral density for full QCD
was evaluated in the microscopic domain of 
full QCD using the method of complex orthogonal polynomials \cite{O}. 
It was found that it  has oscillations in a macroscopic
region of the complex plane  with an amplitude that increases exponentially
with the volume and a period that is inversely proportional to the volume
\cite{AOSV}. The cancellations of the exponential large contributions 
result in a condensate that has a discontinuity  
in the thermodynamic limit \cite{OSV}.

The same phenomenon occurs for QCD in one dimension. Although the
eigenvalues of the Dirac operator are located on an ellipse instead
of being scattered in the complex plane, the mechanism for generating
a discontinuity in the chiral condensate is the same as for QCD. 
To simplify the argument we will restrict ourselves to
giving a detailed discussion
for the case of  QCD in one dimension only.

\section{One-dimensional QCD}

\begin{center}
\begin{figure}[!t]
\includegraphics[width=12cm]{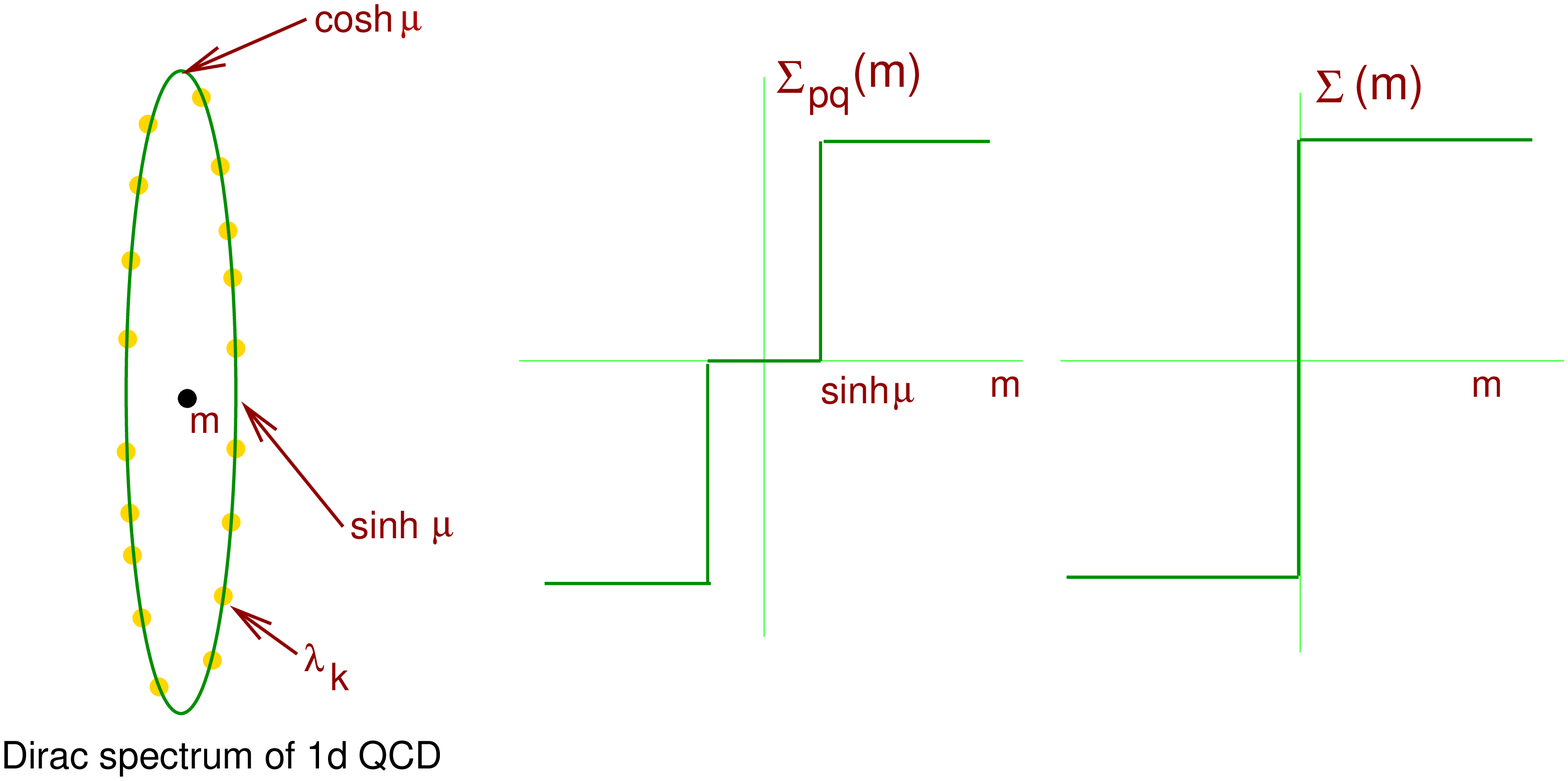}
\caption{The eigenvalue distribution of the one-dimensional QCD
Dirac operator at nonzero chemical potential and large $N_c$ (left). 
The graphs show the mass dependence
of chiral condensate of the partially quenched theory (middle) 
and the full theory (right).}
\label{bl2}
\end{figure}
\end{center}

All gauge fields in one-dimensional QCD can be gauged away with the exception
of the gauge field at the boundary of the manifold. Therefore one-dimensional
lattice QCD is given by the random matrix theory
\be
Z_{1d} = \int dU {\det}^{N_f}D
\ee
with Dirac operator for $n$ lattice points given by.
\be
D = \left ( \begin{array}{cccc} 
mI & e^{\mu}      & {\ldots} &  e^{-\mu}U^{\dagger}\\
-e^{-\mu} & mI & \cdots  &0\\ 
\vdots& &&\vdots\\
 0& \cdots&mI&e^\mu\\
-e^{\mu}{}U/2& \cdots &-e^{-\mu}&mI\\
\end{array} \right).
\label{dirac1d}
\ee
The chemical potential can also be gauged to the boundary.
The full theory
does not have charged excitations resulting in a chiral condensate that
is $\mu$-independent. The phase quenched theory, on the other hand, undergoes
a phase transition at $\mu = m_\pi/2$ resulting in a different free energy.
This can be easily shown in the limit of large $n$ by
 using the explicit result for 
the determinant of the Dirac operator (\ref{dirac1d})
\be
\det D = 2^{-nN_c}\det[ e^{n\mu_c}+e^{-n\mu_c}+e^{n\mu_c}+e^{n\mu_c}U+
e^{-n\mu_c}U^\dagger].
\ee
Here, we introduced the critical chemical potential
$\mu_c$ given by the relation $\sinh\mu_c=m$.
For gauge group $U(N)$ the integrals are particularly simple, 
and we obtain in the limit
of large $n$
\be
F_{N_f = 2} = - n N_f| \mu_c|, \qquad
F_{N_f =2,\,{\rm pq}} =- n N_f |\mu_c|  - n N_f (|\mu|-|\mu_c|)  
\theta(|\mu| -|\mu_c|),\nn\\
\ee
resulting in a chiral condensate with a mass dependence as shown in
Figure \ref{bl2}.
Although nothing happens to the $N_f =2$ free energy, this comes only
as a result of exponentially large cancellations in the partition function.
This is the  ``Silver Blaze Problem'' \cite{cohen} mentioned before. 
We will now 
illustrate how this problem manifests itself in spectrum of the Dirac operator
and the chiral condensate.
 
For simplicity we will only consider the case $N_f=1$ with $U(1)$ as gauge 
group. Because the phase angles of the eigenvalues are uniformly distributed
along an ellipse  with semi-minor axis equal to $\sinh \mu$ and semi-major axis
equal to $\cosh \mu$ (See Figure \ref{bl2}. Notice that 
the figure is for large $N_c$. For $U(1)$ the phase angles of the eigenvalues 
are equally spaced.), 
the eigenvalue density of the $N_f=1$ theory is given by
\be
\rho_{N_f=1}(z)d^2z = \frac 1{2\pi}
\frac{e^{n\mu_c}+e^{-n\mu_c} -e^{n(i\alpha +\mu)}-e^{-n(i\alpha +\mu)}}
{e^{n\mu_c}+e^{-n\mu_c}}\delta(r-\mu)dr d\alpha
\ee
with the original variables $z$  parameterized as
\be
z = \frac 12 (e^{r+i\alpha} - e^{-r-i\alpha}), \quad r >0, \, \alpha \in[0,2\pi].
\ee
The chiral condensate for $N_f = 1$ given by
\be
\Sigma(m)  
=\frac {\left \langle \sum_k \frac 1{\lambda_k+m} 
{\prod_k(\lambda_k+m)}\right \rangle }
 {\left \langle
{\prod_k(\lambda_k+m)}\right \rangle }
\ee
can then be expressed as
\be
\Sigma_{N_f=1} 
&=& \int d^2z \frac {\rho_{N_f=1}(z)}{z+m}\nn\\
&=&\int \frac {d\alpha }{2\pi} 
\frac{e^{n\mu_c}+e^{-n\mu_c} -e^{n(i\alpha +\mu)}-e^{-n(i\alpha +\mu)}}
{(e^{n\mu_c}+e^{-n\mu_c})
(m+(e^{\mu+i\alpha} -e^{-\mu-i\alpha} )/2)} \nn \\
&=& \frac {\tanh(n\mu_c)}{\cosh \mu_c}. 
\ee
When $|\mu| < |\mu_c|$, all four terms in the numerator contribute to the
integrand.
When $|\mu| > |\mu_c|$,
i.e. the domain where the quark mass is inside the ellipse of eigenvalues,
the quenched chiral condensate, given by the first two terms of the 
numerator,  is zero resulting in a chiral condensate that behaves as
the phase quenched theory (see middle panel of Figure \ref{bl2}.
Therefore the condensate for $|\mu|> |\mu_c|$ is due to the oscillating
terms in the numerator. A finite result is obtained because the oscillations
cancel the exponential growth with $n$ of the amplitude. In the thermodynamic
limit the $\tanh \mu_c$ results in a discontinuity at $m=0$ (see Figure
\ref{bl2}).

\section{Conclusions}
Random matrix theory has been invaluable for understanding QCD at
nonzero chemical potential. In this lecture we have discussed applications
involving the phase of the fermion determinant at nonzero chemical 
potential. We have shown that QCD has a severe sign problem if the 
quark mass is inside the domain of eigenvalues. In this domain 
strong cancellations lead to a free energy that does not depend on
the chemical potential and a chiral condensate that has a discontinuity
when the quark mass crosses the imaginary axis. The latter happens
without an accumulation of eigenvalues on the imaginary axis, but
due to oscillations in the spectral density with an amplitude
that increases exponentially with the volume. This mechanism
occurs both in the random matrix limit of QCD and in one dimensional
QCD  which strongly suggests that it is the  generic replacement 
of the Banks-Casher formula for theories with a sign problem.

\vspace{4mm}

\noindent
{\sl Acknowledgments.} 
This work was
supported  by U.S. DOE Grant No. DE-FG-88ER40388 (JV), the 
Carlsberg Foundation (KS), the Villum Kann Rassmussen Foundation (JV) and
the Danish National Bank (JV).

\end{document}